# Atomic structure of CdS magic-size clusters by X-ray absorption spectroscopy

Ying Liu,*[a] Lei Tan,[a] Giannantonio Cibin,[b] Diego Gianolio,[b] Shuo Han,[c] Kui Yu,[c] Martin T. Dove[a,d,e] and Andrei V. Sapelkin[a†]

a. Centre of Condensed Matter and Materials Physics, Queen Mary University of London, Mile End Road, E1 4NS, London, UK

b. Diamond Light Source Ltd., Harwell Science and Innovation Campus, Didcot, OX11 0DE, UK.

c. Institute of Atomic and Molecular physics, Sichuan University, Chengdu People's Republic of China

d. Department of Physics, School of Sciences, Wuhan University of Technology, 205 Luoshi Road, Hongshan district, Wuhan, Hubei, 430070, People's Republic of China

e. College of Computer Science, Sichuan University, Chengdu 610065, People's Republic of China

† Corresponding authors e-mail: a.sapelkin@qmul.ac.uk

Magic-size clusters are ultra-small colloidal semiconductor systems that are intensively studied due to their monodisperse nature and sharp UV-vis absorption peak compared with regular quantum dots. However, the small size of such clusters (<2 nm), and the large surface-to-bulk ratio significantly limit characterisation techniques that can be utilised. Here we demonstrate how a combination of EXAFS and XANES can be used to obtain information about sample stoichiometry and cluster symmetry. Investigating two types of clusters that show sharp UV-vis absorption peaks at 311 nm and 322 nm, we found that both samples possess approximately 2:1 Cd:S ratio and have similar nearest-neighbour structural arrangements. However, both samples demonstrate a significant departure from the tetrahedral structural arrangement, 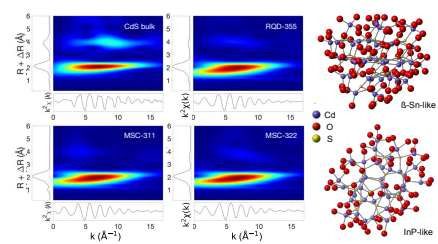 with an average bond angle determined to be around 106.1° showing a bi-fold bond angle distribution. Our results suggest that both samples are quazi-isomers – their core structure has identical chemical composition but a different atomic arrangement with distinct bond angle distributions.

## Introduction

Inorganic quantum dots (QDs) are nanoparticles with dimensions of around 1–100 nm and exhibit significant size-dependent changes in the electronic and optical properties due to the quantum confinement effect[1–8]. Thus, precise control of particle size is one of the critical elements in the synthesis of functional inorganic QDs. Over the past decades, several synthesis methods have been developed for this purpose, including molecular beam epitaxy, metal-organic chemical vapour deposition and colloidal techniques. The colloidal process, in particular, has been studied extensively due to a combination of simplicity and reproducibility[9]. As a consequence, the unique properties of QDs and advances in synthesis have resulted in the wide range of application[10] in multiple areas including bio-imaging[11], quantum LED[12], lasers and electroluminescent devices[13]. Virtually all the above applications rely on the ability to tune emission spectra with, preferably, narrow emission lines. While the emission peak position depends on the particle size and can now be precisely controlled, the emission linewidth is determined, among other things, by the particle size distribution, which is typically finite (i.e. a system is not monodisperse). The latter is a consequence of the classical nucleation and growth process[14].

However, it has recently been found that during the nucleation process, persistent optical absorption peaks appear and remain sharp[15–27]. Their peak position and width suggest that they originate from ultra-small particles of identical size (i.e. monodisperse). Mass spectrometry and optical absorption suggest that these QDs do not follow the continuous nucleation-and-growth model typical for colloidal synthesis. Recently, a two-pathway model was proposed for the development of colloidal QDs [28–30], the model of which contains the evolution of these particles, named magic-size clusters (MSCs) to distinguish them from the conventional ultra-small regular quantum dots (RQDs).

These MSCs provide great promise in atomic-scale control of QDs with electronic and optical properties engineered precisely for applications. Understanding the atomic structure of MSCs is essential for gaining insights into their electronic and optical properties, as well as for the understanding of their synthesis. Besides, it has recently been found that some CdS MSCs of size around 2 nm undergo a reversible isomerisation transition on changing temperature in which one optical absorption peak will gradually disappear, and another peak grows with the two peaks located at 311 and 322 nm[31]. However, the exact nature and the pathway of this transformation are still under debate. Recent work suggests that these two CdS MSCs possess different atomic structures[32], although the nature of the differences is unclear. This

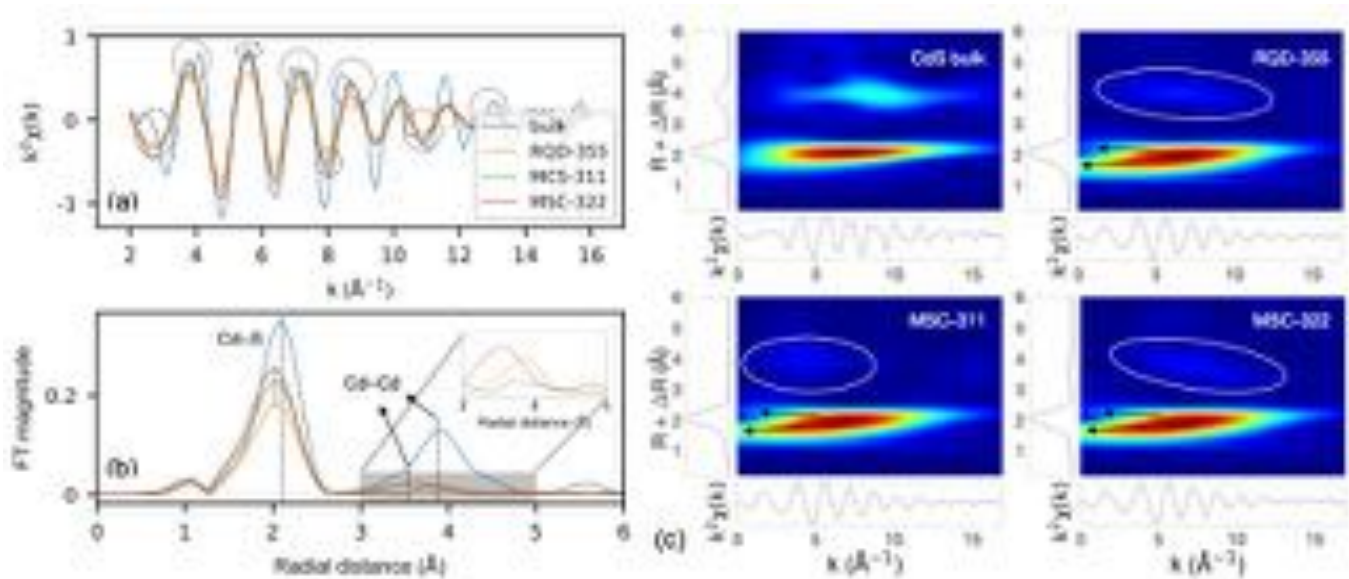

**Fig. 1** (a) The EXAFS oscillation of CdS RQD-355, MSC-311 and MSC-322 samples with bulk reference CdS (space group: $F\bar{4}3m$). The continuous circles label out the difference between bulk reference and QD samples and the dashed circles label out the difference between two MSCs. (b) The magnitude of Fourier transforms (without phase shift) in the corresponding EXAFS in (a). The intensity indicates the possibility of finding atoms at this radial distance from absorbing atoms. QDs show shorter atomic distance than bulk CdS in the first peak. The second peak is amplified in the inset figure to show the difference in the Cd–Cd peak in QDs. (c) Wavelet transforms of corresponding $k^2\chi(k)$ in (a) are shown here. The second $r$-space peak (at about 4 Å) in RQD and MSC-322 samples show a similar wavelet transform feature (in white circle) while the one in MSC-311 is different from the other two.

is a consequence of the significant challenges encountered using standard structural analysis techniques (e.g. X-ray and electron diffraction, Raman scattering, etc.) in the characterisation of ultra-small non-periodic systems. Recent analysis[29] of MSCs' structure using the X-ray pair-distribution function suggests that their atomic arrangements are different from those found in the bulk CdS (zinc-blende or wurtzite structure) and the regular QDs (cubic, zinc-blende structure). This prompted a random structure search approach[33] that indicated a gradual transformation from a cage-like to a bulk-like structure as the number of atoms in a cluster increase. Despite these recent advances in structural characterisation, the atomic structure of the CdS MSCs is still under debate.

Here we utilised X-ray absorption spectroscopy (XAS) to establish a rational basis for understanding the local atomic arrangement and Cd:S ratio in MSC samples. We found that our methodology can yield both local atomic arrangement information (through X-ray absorption near edge structure, XANES), structural and stoichiometry information (through extensive X-ray absorption fine structure, EXAFS) that can be used as an input for subsequent analysis of scattering data and computer modelling.

## Results and discussion

It is previously established[32] that the CdS MSC with 311 nm and 322 nm (labelled as MSC-311 and MSC-322 respectively) optical absorption wavelength have a particle size of about 1–2 nm. Here, we used both EXAFS and XANES parts of the X-ray absorption spectra for the analysis of their atomic structures.

**Extended X-ray absorption fine structure and model-free data analysis**

The recorded EXAFS signals ($k^2\chi(k)$) following background removal and the corresponding magnitudes of Fourier transforms (FT) of CdS samples (MSC-311, MSC-322, RQD-355 and the crystalline reference) are illustrated in Fig. 1a and 1b. Here, $k^2\chi(k)$ is the oscillation observed in the absorption coefficient as a result of photoelectron interference. From the $k^2\chi(k)$ data in Fig. 1a, we can see the difference between QDs signal and the crystalline CdS signal as the later contains pronounced oscillation with high-frequency wave (shown in grey circles) and slower decaying amplitude at high $k$ range, indicating a more ordered structure. The discrepancies between the three QDs are relatively small. Some shifts in oscillation between MSC-311 and MSC-322 can be spotted at 3, 6 and 8 Å$^{-1}$ in Fig. 1a (see dashed circles). This observation is consistent with the X-ray PDF result published previously[32] as MSC-311 and MSC-322 showed different radial distance distribution functions in the long $R$ range (5 – 15 Å) in PDF.

The corresponding FT magnitudes in Fig. 1b reveals the information about the atomic structure around the absorbing atom. Here, we observe differences in the radial distance peak position for the first peak (about 2.1 Å) and the second peak (about 3.8 Å). There is also a small reduction in the interatomic distance in the first FT peak position in the bulk-RQD-MSC sequence. The first peak in bulk CdS at around 2.2 Å corresponds to the Cd–S shell at around 2.2 Å. The first shell in the CdS QDs is broader and shifted to a shorter distance. The signal from the second peak (Cd–Cd shell, features prominently in bulk CdS just below 4 Å) is very weak in RQD and MSCs indicating a significant degree of disorder. The different locations of second peak (3.5 Å vs 3.9 Å, see inset in Fig. 1b) in MSCs compared to the RQD indicate significant bond angle distortions in these systems compared to the bulk crystalline sample. Furthermore, the MSC-311 sample shows no clear second shell (Cd-Cd) signal suggesting that the disorder is the largest for this sample. However, while apparent differences (at about 3, 8 and 10 Å$^{-1}$) can

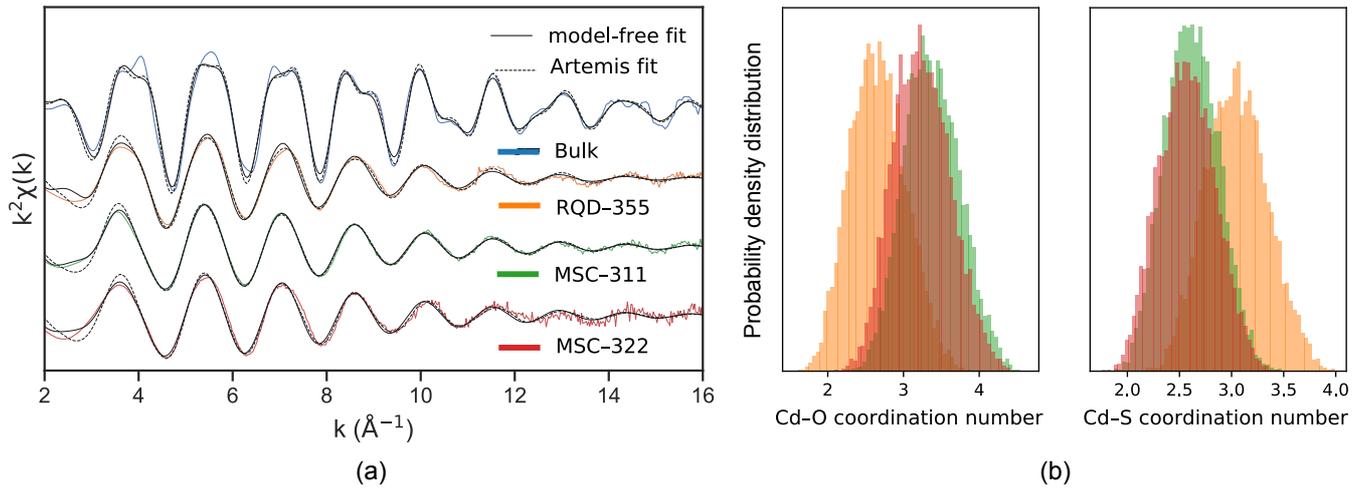

**Fig. 2** (a) The fitted result from model-free fit approach (continuous black lines), the fitted result from Artemis fit approach (dashed black lines) and the experimental EXAFS (coloured lines) for CdS bulk, RQD-355, MSC-311 and MSC-322 samples. The fitted model within energy range 2–16 Å$^{-1}$ for all experiments is the sum of three single scattering paths: Cd–O, Cd–S and Cd–Cd. Energy dependant parameters, backscattering amplitude $f_j(k)$ term and phase shift $\delta_j(k)$, were extracted from bulk experimental data in model-free fit or calculated using FEFF package in Artemis fit. The radial distance of Cd–O and Cd–S is set like the ones in bulk CdO and bulk CdS. (b) The probability density distribution of fitted Cd–O and Cd–S coordination number for three QDs. Each distribution is the histogram of the fitted result after 10,000 fits. The labelled colour for each QD is the same as the ones in (a). In this figure, we can see that MSC-311 and MSC-322 has a large overlapping area in coordination number distribution.

be observed between RQD and MSCs in the $k^2\chi(k)$ data, these are not as obvious in the corresponding FTs that look somewhat similar, especially the first peak. Hence, we used Cauchy wavelet transform (see Fig. 1c) which is capable of visualising EXAFS spectra in three dimensions: the wavevector $k$, the radial distance $R$ and wavelet transform modulus. As a result, we determined multiple contributions to the first radial distance peak. Indeed, a split (indicated with arrows) was observed in the first peak originating from low $k$ range. It shows a substantial contribution to the overall shape from two sources. One is at slightly shorter $R$ (around 1.8 Å in Fig. 1b), and another one is at a longer $R$ (about 2.2 Å in Fig. 1b) suggesting the presence of Cd-O bonding in the sample. Furthermore, on close inspection, the second peaks (at about 4.0 Å) in RQD-355 and MSC-322 shown in wavelet transformation have similar patterns with a $k$-space contribution between 3 Å$^{-1}$ and 12 Å$^{-1}$. In contrast, in MSC-311, most of the input comes from the 2–8 Å$^{-1}$ range. This is a clear indication of structural differences not only between the bulk CdS and nanoparticles but also between the two MSCs.

Having identified detectable differences in EXAFS signal between MSCs, we carried out a numerical analysis to further explore their nature. Our approach to EXAFS analysis is based on the modern X-ray absorption spectroscopy theory that a combination of experiments together with the data analysis can yield an agreement between the data and a model down to the noise level[34]. The EXAFS signal $\chi(k)$ can be described by the following expression[35].

$$\chi(k) = \sum_j \frac{N_j S_0^2 f_j(k)}{k R_j^2} e^{-2\sigma_j^2 k^2} e^{2R_j/\lambda_j(k)} \sin\left(2ik \cdot R_j + i\delta_j(k)\right) \quad (1)$$

where $\chi(k)$ is a sum over unique scattering paths $j$ of the photoelectron, $f_j(k)$ is backscattering amplitude, $\delta_j(k)$ is phase shift, $\lambda_j(k)$ is the inelastic mean free path, $S_0^2$ is many-body amplitude reduction factor accounting for the effects of inelastic losses, $N_j$ is the coordination number, $R_j$ is the half of scattering path length (inter-atomic distance in the case of single scattering), $\sigma_j^2$ is mean squared atomic displacement relative to the absorbing atom (EXAFS Debye-Waller factor). Some of the parameters ($f_j(k)$, $\delta_j(k)$, $\lambda_j(k)$) are generally calculated ab initio using multiple scattering (MS) software such as FEFF[36] while others are used as fitting variables: $N_j S_0^2$, $\sigma_j^2$ and $R_j$.

In the widely used EXAFS analysis package DEMETER[37], the oscillatory EXAFS signal is fitted by least-squares routine, but the analysis procedure requires a structural model as an initial guess in the fitting process. Such an approach works well for systems where a reasonable assumption can be made about the structure based on complementary information obtained from other sources. However, the problem with MSCs is that their structure is generally unknown. The bulk CdS structure may be a misleading initial guess since for small clusters a variety of structures can be stable (or metastable) depending on particle size[33].

Hence, we carried out further refinement using a custom-written Python code minimising the sum of squared residual[38] between fit result and experiment, while making no assumptions about the structure of the sample. Such an approach ignores MS calculations, but the MS provides a negligible contribution to the overall signal in low-coordinated systems with large disorder[34]. In this model-free analysis, we treated the EXAFS signal of a single path as amplitude and frequency-modulated function of the form

$$\chi_j(k) = A_j(k) \sin\left(2kR_j + \delta_j(k)\right) \quad (2)$$

with the amplitude $A_j(k)$

$$A_j(k) = S_0^2 \frac{N_j f_j(k)}{k R_j^2} e^{-2\sigma_j^2 k^2} e^{2R_j/\lambda_j(k)} \qquad (3)$$

We considered three scattering paths to fit the EXAFS equation: Cd–O, Cd–S and Cd–Cd (second shell). Each path can be represented individually by eqn (1). The sum of three paths is the $\chi(k)$ which corresponds to the experimental EXAFS signal. The amplitude term $f_j(k)$ and phase $\delta_j(k)$ have been extracted from eqn (2) and (3) using bulk CdS and CdO data with the procedure similar to that described previously[39]. The corresponding structural parameters in the bulk references are listed in Table S1. We compared extracted and calculated (using FEFF) scattering amplitudes and phases for the three scattering paths (see Fig. S1). While the amplitude and phase are similar in Cd–O path, the amplitudes in the other two paths are larger at low $k$ range and smaller at high $k$ range in the experimental data than the calculated ones. In the following analysis, we used parameters extracted from the experimental bulk reference to fit the model (eqn. 1) to the experimental data. The fitted parameters are $N_j S_0^2$ and $\sigma_j^2$ in three scattering paths: Cd–O, Cd–S and Cd–Cd, while $R_j$ were defined based on the results of PDF analysis[29]. The fitting process was carried out using a differential evolution algorithm[40] to find an optimum of the minimum of the sum of squared residuals (SSR) between experiment and model. The SSR value was used to compare with the fitting the widely used Artemis[37] EXAFS analysis code. The fitted spectra for three QDs and the bulk reference are shown in Fig. 2a where the positions and amplitudes of fitted curves in both model-free approach and Artemis are in good agreement with experimental data. We chose the model-free fit result rather than Artemis here since it utilised backscattering amplitudes extracted from the experimental reference. The probability density histograms for fitted coordination numbers are shown in Fig. 2b. Fitted variables for the model-free approach are listed in Table 1. The value of the amplitude factor, $S_0^2$, is typically to be close to 1[41]. Thus, $N_j S_0^2$ can serve as a reference for relative changes in $N_j$. Numerical analysis shows both MSCs have a similar coordination number while RQD-355 shows a smaller Cd–O and larger Cd–S coordination number. The results suggest a reduction in the first shell Cd-S coordination numbers in the CdS bulk–RQDs–MSC sequence, indicating a smaller particle size for MSCs. This is consistent with the PDF data reported previously[29]. At the same time, the Cd–O coordination numbers show a slight increase, which is again consistent with the smaller particle size for MSCs and, therefore, a larger surface-to-volume ratio of Cd atoms. Furthermore, observation of the second coordination shell (Cd–Cd) in the FT magnitude MSC-322 (see inset in Fig. 1b) provides an opportunity to calculate the average value of Cd-S-Cd bond angle with the fitted Cd–S distance of 2.4881(98) Å and Cd–Cd distance of 3.9769(96) Å. This was found to be 106.1(1.1)°, which is lower than 109.47° expected for a perfect tetrahedral structure. The data also suggest that angle disorder is larger in MSC-311 sample since no clear second shell signal can be seen.

Table1. The structure parameters of three QDs from the structure model-free fitting approach using eqn (1).

|  | scattering path | $N_j S_0^2$ (mean) | Standard deviation | SSR Model-free | Artemis |
|---|---|---|---|---|---|
| RQD-355 | Cd-O | 2.67 | 0.37 | 0.85 | 0.64 |
|  | Cd-S | 3.05 | 0.30 |  |  |
|  | Cd-Cd | 5.76 | 3.31 |  |  |
| MSC-311 | Cd-O | 3.38 | 0.37 | 0.22 | 0.47 |
|  | Cd-S | 2.62 | 0.26 |  |  |
|  | Cd-Cd | 5.57 | 3.25 |  |  |
| MSC-322 | Cd-O | 3.24 | 0.38 | 1.08 | 1.24 |
|  | Cd-S | 2.57 | 0.27 |  |  |
|  | Cd-Cd | 5.40 | 3.28 |  |  |

The data in Table 1 clearly show that Cd-S coordination is lower than the value found in the bulk CdS (4). This, together with the increase in Cd-O coordination, points to the large proportion of Cd atoms located at the surface of QDs. Based on these observations, with the mass spectroscopy results[31] of MSC-311 and MSC-322 MSCs as ~5160 Da in atomic weight, we can calculate the stoichiometry of $Cd_xS_y$. The calculation is based on two assumptions: i) MSCs remain four coordinated; ii) S atoms are fully capped with Cd. Since MSCs are monodispersed particles, we defined both $x$ and $y$ as integers. According to the mass spectroscopy result, $Cd_xS_y$ has an atomic mass around 5160 Da. Thus, we listed all possible combination which has an atomic mass around 5160. The Cd–S coordination number in MSC-311 and MSC-322 is around 2.6. Based on assumptions, the coordination number of Cd–S in $Cd_xS_y$ is calculated as $4y/x$. We can then limit our choices to $Cd_{38}S_{28}$, $Cd_{39}S_{24}$ and $Cd_{40}S_{21}$. These combinations have Cd–S coordination number of 2.95, 2.46 and 2.10, and atomic masses are 5169, 5154 and 5170 Da respectively. The number of oxygens on the cluster surface depends on their surface Cd number. Among the three option, $Cd_{39}S_{24}$ has the coordination number which sits within closest to the MSC data in Table 1, suggesting that the stoichiometric ratio is $x/y \approx 2$.

**X-ray absorption near-edge structure**

Having analysed EXAFS data, we turned our attention to the near edge part of the spectra that is sensitive to the oxidation states and to the local symmetry. XANES data (K and L3 absorption edges) for bulk CdS, MSCs and their derivatives are shown in Fig. 3a and 3b. In Fig. 3a, all QD spectra show sharp absorption ("white line") just above the absorption K-edge, while retaining some of the features of the bulk CdS spectrum. The sharp white lines at Cd K-edge XANES of these QDs suggest the presence of oxygen around Cd atoms. The data also indicate that there is a gradual evolution of XANES as the cluster size is reduced (from bulk to RQD to MSC, see Fig. S2 in ESI), which is particularly

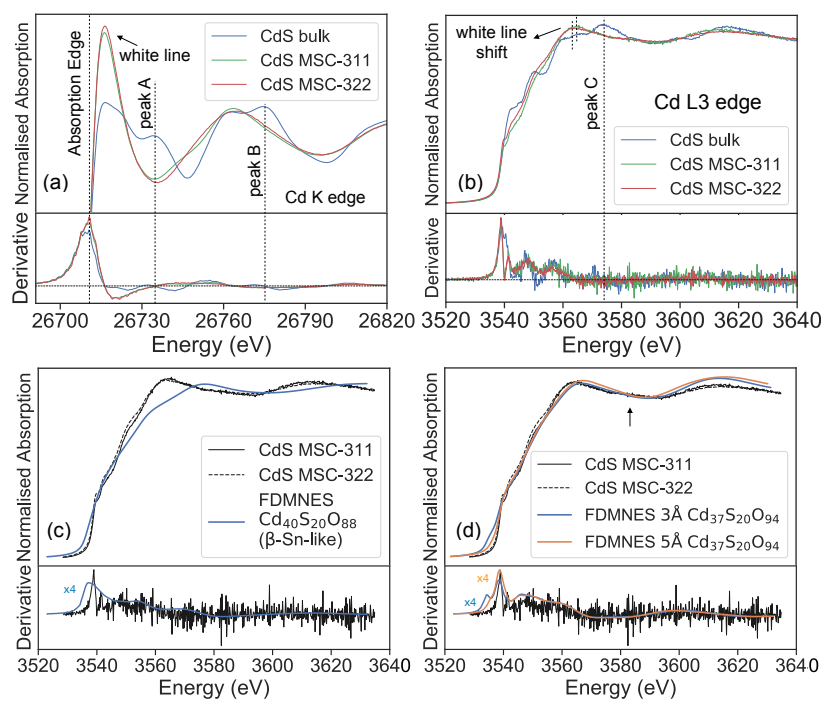

**Fig. 3** (a) Cd K edge XANES and their first derivative of CdS MSC-311 and MSC-322 samples with bulk reference CdS. The maximum first derivative is defined as the absorption edge. The peak A and B from bulk XANES are not observed in MSC spectra. (b) Cd L3 edge XANES and their first derivative of the sample in (a). MSC-322 spectrum is observed with a small shift in white line from that in MSC-311. (c) FDMNES simulated CdS cluster with the core of β-Sn-like structure and surface Cd terminated with O. (d) FDMNES simulated CdS cluster with the core of distorted zinc-blende structure (InP-like) and surface Cd terminated with O.

evident for peak B. Still, evolution can also be seen for the peak A. These data suggest significant changes in the local atomic environment and/or local coordination. To shed light onto the origins of these changes, we investigated L3-edge absorption spectra where the energy resolution is better than at the K-edge due to the increased core-hole lifetime. Here again, a difference in 3530 eV and 3590 eV range can be most clearly observed between the bulk sample and MSCs (Fig. 3b). The most obvious result here is the disappearance of the feature designated as peak C and development of a prominent "white line" at around 3560 eV, while the gradual evolution of the spectra below 3560 eV suggesting the increase of a local disorder without significant structural changes.

To further explore the nature of the observed difference, we carried out XANES calculations utilising the FDMNES code[42] and stoichiometric information we obtained from the EXAFS data. For this purpose, several structures were prepared, including crystalline systems and clusters. It is well-known that reduction of system size down to the nanoscale results in an increase of the surface-to-bulk ratio of atoms, inevitably leading to a more significant influence of surface oxides[43,44] and contraction of the average bond length[45,46] at the surface. The reduced bond length results in higher local electron density. Increased electron density is also found in materials under pressure[47,48]. Therefore, it is logical to assume that these systems may adopt structures favoured under compression. Hence, we investigated the effects of (i) oxidation, (ii) structure (i.e. polymorphism) and (iii) size (e.g. bulk vs nanoscale) on XANES spectra. Within this context, model systems with tetrahedral ($F\bar{4}3m$, zinc-blende), and octahedral ($Fm\bar{3}m$) structures were investigated. Furthermore, it is known that the transformation path under compression in ultra-small semiconductor nanoparticles with the tetrahedral structure results in the octahedral-like distortion towards the β-Sn ($I4_1/amd$) structure[49]. Hence, this structure has also been included in our simulations.

We first examined the effect of oxidation. From the calculated results of CdO as a function of coordination shell number around absorbing atom (see Fig. S3), we quickly established that CdO is responsible for the sharp white line in the MSC XANES at around 3564 eV. We also established that the contribution of CdO in MSCs does not extend beyond the first coordination shell (see the evolution of the CdO XANES in Fig. S3). Based on the simulated CdO spectra, the slight difference in the intensity of the white line and a small energy shift (about 1.2 eV) between for MSC-311 and MSC-322 samples (indicated with the arrow in Fig. 3b) is due to the different oxygen content in MSC samples. Derivatives of the experimental spectra and CdO single-shell model indicate that the latter reproduces some, but not all the features in the data. This is not surprising as we already know from EXAFS analysis that Cd-S bonding is also present in all samples.

We then investigated the possibility for several polymorphs of CdS to be responsible for the observed XANES signal. To that end, we attempted to reproduce the experimental data as a weighted sum of CdS and CdO with signals calculated within 3 Å around the absorbing atoms (Cd) from their corresponding crystal structures. We found that neither tetrahedral ($F\bar{4}3m$, zinc-blende), nor octahedral ($Fm\bar{3}m$) structures could adequately reproduce the experimental data (see Figs. S4 and S5) with the β-Sn-like ($I4_1/amd$) arrangement providing closest fit (Fig. S6). These results suggest that there is some sort of systematic distortion of the tetrahedral structure (of which β-Sn-like arrangement can be considered a limiting case[49]) in these systems.

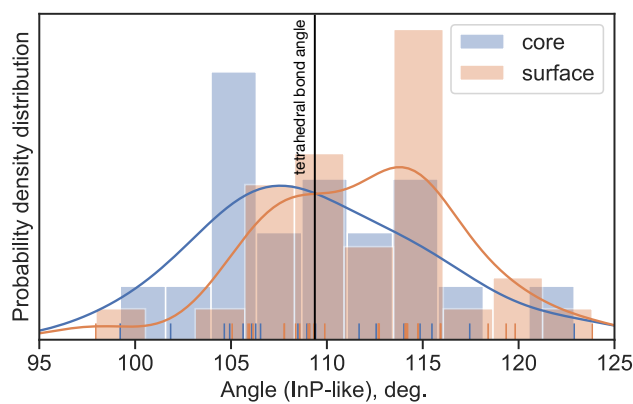

**Fig. 4** The bond angle distribution histogram around the core atoms and surface atoms in CdS model with InP-like structure. Core label refers to the angles around the atoms which are bond with only Cd or S. Surface label refers to the angles around the atoms which are bond with O. The curves are the gaussian kernel density estimate for each histogram and tick marks sit at the values of bond angles.

Having identified the β-Sn-like structure as a potential candidate, we prepared an oxygen-terminated cluster of a suitable size (based on the mass spectroscopy results[31]) and our EXAFS results above. We also introduced another candidate: the InP-like[32,50] structure that has been suggested previously to be a good atomistic model both for MSC-322 and MSC-311 clusters[34] while its stoichiometry ($In_{37}P_{20}$) fits well with our EXAFS data. Again, an oxygen-terminated cluster of suitable size has been prepared for XANES calculations. The results of XANES calculations are shown in Fig. 3c, d and demonstrate excellent agreement of the $Cd_{37}S_{20}O_{94}$ InP-like signal with the experimental data (Fig 3d). This is consistent with the X-ray scattering results reported previously[32]. However, there is still a small discrepancy at around 3585 eV (see arrow in Fig 3d). This discrepancy can be accounted for by the white line at about 3580 eV in the β-Sn-like cluster (Fig. 3c), thus suggesting that a related distortion may be present.

Considering the results above, we analysed the bond angle distribution in the InP-like CdS cluster used for XANES calculations. The results are shown in Fig. 4, together with the corresponding angles of the perfect tetrahedral. Angle distribution analysis suggests that there is a tendency for bi-fold angle distribution with the maximum of the angle distribution in the core at around 107°, while the corresponding maximum at the surface is around 115°. The average bond angle for the core is in good agreement with the EXAFS data reported above where the value of 106.1(1.1)° was found. Therefore, we conclude that the introduction of bi-fold distortion into the perfect tetrahedral system results in a better agreement between calculations and experimental data and it is among such systems that one should be looking for potential candidates for a model structure.

## Conclusion

The XAS data were analysed for two CdS MSCs samples prepared by the colloidal synthesis method. EXAFS analysis showed a reduced first coordination shell distance and a broader distance distribution around Cd atoms in MSCs than in the bulk reference. The split in the peak corresponding to the first coordination shell observed in the wavelet transformation indicated that Cd atoms are also bonded with another element besides S atoms. Both XANES and EXAFS show that the surface of CdS clusters is capped with O bonded to Cd. We used a model-free structural analysis method to establish that the stoichiometric Cd:S ratio of the MSCs is 2:1. XANES analysis revealed that out of several potential candidates examined, the InP-like structure reported previously[32] results in the best fit to the data. We also show that the average bond angle in the core of CdS clusters is 106.1(1.1)° and is well below the value 109.47° expected for a perfect tetrahedral structure. Further analysis of the bond angle distribution in the InP-like cluster used for the best fit to the XANES signal shows evidence of bi-fold bond-angle distribution between the core and the surface of the CdS MSCs. Finally, combined EXAFS and XANES analysis suggests that it is the change in the bond angle distribution that is responsible for the recently observed[31] thermally-induced reversible structural isomerisation in these systems.

Overall, we conclude that combination of EXAFS and XANES analysis can be used as an effective tool to investigate the structure of MSCs as it allows to establish sample stoichiometry and test potential structural models against the experimental data. These data can then be used as effective guidance to construct atomistic models of MSCs.

## Experiment

Two CdS MSC samples and one RQD sample with UV-vis absorption peaks at 311 nm (MSC), 322 nm (MSC) and 355 nm (RQD) were analysed here together with the CdS bulk reference. They are labelled after QDs type and absorption peak position as MSC-311, MSC-322 and RQD-

355. Their absorption spectra are shown in Fig. S7. MSC-311 could transform to MSC-322 after heating at 50 °C while MSC-322 could transform back at a temperature lower than 25 °C. All the QDs sample are synthesis in oleic acid resulting in oleic acid capped products. XAS measurement (performed at beamline B18, step-size: 0.4 eV, energy resolution: $\Delta E/E = 1.4 \times 10^{-4}$) was carried out at the Diamond synchrotron light source. EXAFS was recorded at the liquid $N_2$ temperature (90 K) for Cd K edge and at room temperature for Cd L3 edge. The CdS bulk sample with zinc-blende structure (space group: $F\bar{4}3m$) was used here as the reference to our experimental data.

The XAS data reduction and analysis were carried out using the DEMETER package[37] and Pyspline[51] in the standard way[39]. The parameters used in the data reduction included the background removal threshold $R_{bkg}$ (0.8), the k-window for FT (2–16.5Å$^{-1}$) and the k-weight ($k^2$). The wavelet transformation was calculated with $k^2\chi(k)$ data from previous data reduction process. It used the code from Munoz's group[52]. Then we carried out the data analysis process in two routes: calculation and experiment. In the calculation route, parameters such as backscattering amplitude $f(k)$, mean free path $\lambda(k)$ and phase shift $\delta(k)$ were calculated with the bulk structure using FEFF[36,53]. These parameters were then applied in DEMETER package[37] to fit QD samples. QD samples were also fitted using in-house code in the experiment route. We filtered the EXAFS signal for their first Cd-S, Cd-Cd (in bulk CdS) and Cd-O (in bulk CdO) radial distance peak. Then the $f(k)$ related term and $\delta(k)$ corresponding to each distance peak were computed. In the fitting process, we used the differential evolution method to achieve the global minimum of the SSR between experiment and model for multiple variables. The fitting process was looped 10000 times to create probability density distribution for each variable.

The XANES simulated spectra were calculated using FDMNES[42]. The finite difference method (FDM) was used for XANES calculations, which goes beyond the muffin-tin approximation typically used for multiple-scattering calculation for solving the excited state. CdS zinc-blende crystal structure was calculated to verify the accuracy of FDMNES (see Fig. S8 in ESI). All clusters were relaxed using CrystalMaker[54] prior to XANES calculations.

## Conflicts of interest

The authors declare no conflicts of interest.

## Acknowledgements

YL and LT received financial support from the Chinese Scholarship Council and the Queen Mary University of London. This research utilised Queen Mary's Apocrita HPC facility, supported by QMUL Research-IT[55]. This work has been supported in part by BBSRC (grant number BB/J001473/1)

## Notes and references